\documentclass[11pt,twoside]{article}

\usepackage{asp2006}
\usepackage{lscape}

\begin{document}
\title{Electrons re-acceleration at the footpoints of Solar Flares}   %%% Fill in title
\author{Rim Turkmani}   %%% Fill in author names
\affil{  Department of Physics, Imperial College, London SW7 2AZ}    %%% Fill in author affiliations
\author{John Brown}   %%% Fill in author names
\affil{ Department of Physics \& Astronomy, University of Glasgow, G12 8QQ, UK} 

\begin{abstract} %%% Abstract to run on from here.
 Hinode's  observations revealed a very dynamic and complex chromosphere. This require revisiting the assumption that the chromospheric footpoints of solar flares are areas where accelerated particles only lose energy due to collisions. Traditionally electrons are thought to be accelerated in the coronal part of the loop, then travel to the footpoints where they lose their energy and radiate the observed Hard X-ray. Increasing observational evidence challenges this assumption. We review the evidence against this assumption and present the new Local Re-acceleration Thick Target Model (LRTTM) where at the footpoints  electrons receive a boost of re-acceleration in addition to the usual collisional loses. Such model may offer an alternative to the ÔstandardÕ collisional thick target injection model (TTM) (Brown 1971) of solar HXR burst sources, requiring far fewer electrons and solving some recent problems with the TTM interpretation. We look at the different scenarios
 which could lead to such re-acceleration and present numerical results from one of them. 
\end{abstract}

\section{Introduction}
Our image of the chromosphere has evolved in the last few decades from a simple
 stationary structured layer to a highly dynamic one 
which hosts a very complex magnetic field structure. 
Hinode's high resolution observations surprised us with an image of the
chromosphere that is even more complex than predicted  
(e.g. Liu {\it et al.}  2009) and revived the debate about the relationship between chromospheric
activities and the corona.  This is particularly important In the context of the relationship between the chromospheric footpoints 
of a flaring loop and its coronal part, particularly during the hight of flares which are accompanied by strong HXR radiation at the footpoints. 
The standard model of solar flares view the footpoints as a thick region where particles accelerated at the corona lose their energy due to collisions and radiate HXR. The relationship between the population of accelerated particles and the HXR at the footpoints is given by the Thick Target Model (Brown 71, Hudson 72)
As with other 'Standard Models', steady improvements in relevant data are necessary.
Here we recap the TTM in its basic form with injection from the corona,
indicate some of the issues facing this, and discuss an alternative
type of scenario in which the acceleration and radiation ('target') regions are not distinct.

\section{Assumptions and Problems of the TTM}

One of the main assumptions of Thick Target Model which we question here is the strict
separation between the acceleration and the radiation regions.  
Breaking this assumption and allowing particles to gain energy in the radiation region solves 
may of the challenges which TTM faces as discussed bellow. 
In the TTM fast electrons are assumed to be accelerated (so the net
energy change rate $\dot E >0$) by a usually unspecified process 
in a tenuous coronal region ${\cal A}$ and
injected continuously into a cool dense target chromosphere
radiation region ${\cal R}$ where their energies are assumed to be
controlled mainly by collisional losses (so that $\dot E < 0$)
accompanied by HXR bremsstrahlung radiation. Radiation in ${\cal A}$
is assumed to be small at high energies $\ge$ 20 keV (except for
unusually dense loops - Veronig and Brown 2004). In reality,
non-collisional $\dot E$ due to return current Ohmic dissipation is
important for the more intense events (Brown and Bingham 1984) while some redistribution
of $E$ by wave-particle interactions is inevitable since dispersive
propagation ( Melni'k {\it et al.} 1999, Kontar 2001)
and even collisions alone create electron distributions with
$f'(v)>0$ (Haydock {\it et al.} 2001).
Even in the purely collisional case of TTM injection, observed HXR
fluxes place strong requirements on electron beam density and power
supply. First, in {\it any} cool HXR source, long range Coulomb
energy losses from fast electrons greatly exceed ($\simeq
10^5~\times)$ the HXR power radiated in close collisions, so the
electron power supply needed is $\simeq 10^5~\times$ the HXR
luminosity. Second, since the electrons in ${\cal R}$ are
decelerated monotonically hence 'lost', they have to be continuously
replenished by injection from ${\cal A}$ at a high rate ${\cal F}_1
$s$^{-1}$ above $E=E_1$. In a large HXR event  ${\cal F}_1 \gg
10^{36}$s$^{-1}$ for $E_1=20$ keV (Hoyng {\it et al.} 1976) which, for event duration
$\tau_o\simeq10^3$s, means a total ${\cal N}\gg 10^{39}$
electrons processed. This exceeds the total electrons in a large
dense coronal loop ($10^{27}$cm$^3\times 10^{11}$cm$^{-3}$), a fact
often termed the {\it Electron Number Problem} though the solution
to this was established long ago (Hoyng {\it et al.}1976, Knight and
Sturrock 1977, Colgate 1978). Beam electrodynamic theory 
 shows that the  E-field created by beam injection drives a
return current in the plasma which delivers electrons back to the
corona at rate ${\cal F}_1$, though there are issues over how the
currents close at ${\cal A}$ (Benka and Holman 1994). However, there
{\it is} a real problem with the {\it instantaneous local beam
density} if it has as small a cross sectional area $A$ as the recent
estimates of the impulsive WLF, optical and UK kernel sizes, $<3''$
(Hudson, Metcalf and Wolfson 2006) or $A<10^{16}$cm$^2$. Since
${\cal F}_1\simeq An_1v_1\approx 10^{36}A_{16}n_{10}$s$^{-1}$ where
$A_{16}=10^{-16}A,n_{10}=10^{-10}n$ and $v_1\simeq
10^{10}$cms$^{-1}$, large event values of
 ${\cal F}_1 \gg 10^{36}$ imply a local instantaneous ({\it not} time integrated){\it beam} density
 above $E_1=20$ keV of at least $n_1>10^{10}$cm$^{-3}$. If the beam is accelerated and propagates
 in the corona, this is highly implausible as it equals (or exceeds) the total
 {\it plasma} density in all but the densest loops ever seen (Veronig and
 Brown 2004).  If the 'beam' footpoints are not monolithic
 but comprise a moving filament of even smaller instantaneous area  (Fletcher {\it et al.} 2004)
 then the  density problem would be even worse, even though such morphology
seems more consistent with the values and evolution of soft X-ray
emission measure and temperature (Stoiser, Brown and Veronig 2008).
Note, however, that this density problem of the TTM only arises in
the case of injection from the tenuous corona into the chromsophere.
It is not an issue for thick target HXR production by a beam
accelerated and propagating wholly in dense layers 

A second challenge to the TTM from recent data concerns the
 beam anisotropy. Electrons descending a loop produce HXR bremsstrahlung
 which is quite strongly beamed downwards (e.g. Brown 1972) unless
they are near to isotropic when injected.  Kontar and Brown (2006)
recently emphasized (cf. Langer and Petrosian 1987)  that, because of the
large HXR albedo of the photosphere, such beaming should result in a
HXR spectral 'bump' signature of the albedo around the 30-60 keV
range. Such a feature is observed but only at the level expected for
a near isotropic primary HXR source. Quantitative inference of beam
anisotropy from HXR spectra, both on a few individual events (Kontar
and Brown 2006) using the photosphere as a 'mirror', and
statistically (Kasparova, Kontar and Brown 2007) points to near
isotropy of the electrons in the HXR source. It should be noted,
however, that their methods measure anisotropy with very low angular
resolution - essentially in one upward and one downward direction -
and could be consistent with many distributions having nearly equal
up and down fluxes, though this is far from the usual TTM
assumption. Another result which, if confirmed, would indicate small
anisotropy of fast electrons is the reported absence of detectable
polarization in impulsive H$_\alpha$ kernels (Bianda {\it et al.}
2005) though this has been refuted by Henoux (personal
communication). A nearly isotropic electron distribution does not of
itself preclude the general TTM type of scenario but fast
precipitation of such electrons in any realistic field requires some
non-collisional transport process - e.g. electron cyclotron masering
(Melrose and Dulk 1982) to defeat the effects of magnetic trapping
and so allow fast electron precipitation.
The difference between interplanetary and HXR source electron spectral 
indices has also been shown to be is inconsistent
with the CTTM scenario (Krucker et al. 2007, 2009).

\section{Local Re-acceleration Thick Target Model and Re-acceleration mechanisms}

In the light of these problems with the TTM, we have considered alternatives
(Brown {\it et al.}  2009) to the injection scenario where electrons are injected as in the TTM but constantly reaccelerated locally inside a dense 
HXR source. In these {\it Local re-acceleration thick target models - LRTTM}  the fractional density, and possibly the anisotropy, of the fast electrons are reduced compared to TTM models involving purely collisional transport after injection from the corona. The reason is basically that the photon yield $\zeta$ from one electron during its lifetime is roughly $n_pvQ\tau$ where $Q$ is the relevant bremsstrahlung cross section for the energy range considered and $\tau$ is the electron lifetime. More precisely it is $\int_t n_pvQdt$ which, for each time segment $\Delta t$ of monotonic $E(t)$, can be written $\int_E n_pvQdE/|\dot E|$. Re-acceleration can reduce $\dot E$ below the collisional energy loss rate $\dot E_{coll}$ and/or increase $E$ and hence $\tau$ hence
$\zeta$ to $\gg \zeta_{CTTM}$. This reduces the ${\cal F}_1$ required to yield a specified HXR event flux.
Many of the acceleration mechanisms which work in the corona could be a candidate for 
particle acceleration in the chromosphere. Most of these mechanisms rely on a generated electric field to transfer energy to the particle.
With the chromosphere being so dynamic, the presence of such electric field is almost inevitable \citep{judge06}, the question is whether it is strong
enough to be able to accelerate particles. The electric field needed to  accelerate a particle from 
its thermal energy is the Driecer field, but if the particle already possesses energy above thermal then 
the field needed to accelerate the particle drops as $1/v^2$. This means that mush a smaller electric field
is needed to give energy to particles arriving from the corona with an initial kick in energy. 
This is why the case of re-acceleration in the chromosphere is stronger than the acceleration 
(which is not ruled out here).
Examples of mechanisms which generate electric field that could re-accelerate electrons are
the  currents resulting from the interaction between the emerging flux and the preexisting magnetic field  \citep{tor09}, electric currents generated by Alfv\'en 
waves  \citep{gu08}, and currents from 
reconnection events \citep{tor09}. Also, density gradience has been shown to be responsible of generating electric field 
both in the corona and the chromosphere \citep{vp06} and \citep{vp09}. 
In \cite{brown09}  current sheet cascade at the chromosphere was suggested as one possible LRTTM, 
it is shown by 3D numerical simulations that these currents are able to reaccelerate significant percentage of electrons and 
subsequently increase their life time in the chromosphere and their photon yield.

\end{document}